\documentstyle[12pt]{article}
\begin {document}
\begin{titlepage}
March 1992 \hfill HU Berlin--IEP--92/1
\vspace{6ex}
\Large
\begin {center}
\bf{String amplitudes in arbitrary dimensions}
\end {center}
\large
\vspace{3ex}
\begin{center}
Stefan F\"orste\footnote{email: foerste@convex.ifh.de}
\end{center}
\normalsize
\it
\vspace{3ex}
\begin{center}
Fachbereich Physik der Humboldt--Universit\"at \\
Institut f\"ur Elementarteilchenphysik \\
Invalidenstra\ss e 110, O--1040 Berlin
\end{center}
\vspace{6 ex }
\rm
\begin{center}
\bf{Abstract}
\end{center}
We calculate gravitational dressed tachyon correlators in non critical
dimensions. The 2D gravity part of the theory is constrained to constant
curvature. Then scaling dimensions of gravitational dressed
vertex operators are equal to their bare conformal dimensions.
Considering the model as $ d+2 $ dimensional critical string
we calculate poles of generalized Shapiro--Virasoro amplitudes.

\end {titlepage}

\section{Introduction}
It is the idea of string theory that all elementary particles and their
resonances can be considered as excitations of strings moving through
space time. The main problem of such a picture is that the space time
dimension $d$ has to be 26 (bosonic case) or 10 (super symmetric case)
in order to get results not depending on the metric of the world sheet
swept out by the string.

Then it was proposed in \cite{1} to integrate over all world sheet
metrics which can be considered as an inclusion of 2D quantum gravity.
Unfortunately that approach provides complex scaling dimensions in the
interesting region of space time dimensions between one and 25
\cite{2,3,4,5,6}.

A natural way to overcome that problem is to weight the integration
over all metrics by a factor
$$ \int \mbox{$\cal{D}$} g \longrightarrow \int \mbox{$\cal{D}$} g
e^{-S_{pg}}, $$
where $S_{pg}$ is an action for pure 2D gravity. The Einstein--Hilbert
action does not solve the problem because it is a topological density
in two dimensions. However, there is another natural candidate for
$S_{pg}$ proposed in \cite{7} and rederived in a topological
framework in \cite{8},
\begin{equation}
S_{pg}=\frac{i}{\pi }\int d^{2}z \, \sqrt{g} \phi (R+\Lambda ),
\end{equation}
where $\phi $ is a two dimensional dilaton which has to be taken into
account as a quantum field. $R=R(g)$ is the two dimensional
curvature and $\Lambda $ is a cosmological constant.
(We use conventions of \cite{9}). The dilaton enters the action
like a Lagrange multiplier and hence (1.1) produces the constraint of
constant curvature. There are also possible generalizations of (1.1)
discussed in \cite{10}.

In \cite{11} it has been shown that the inclusion of (1.1) provides a real
string susceptibility (scaling dimension of the partition function)
in arbitrary space time dimensions $d$. Later on it was shown in \cite{12}
(perturbative approach) and in \cite{13} (non perturbative approach) that
scaling dimensions of vertex operators are equal to their bare conformal
dimensions and hence real. (The perturbative approach is also discussed
in \cite{14}).

The mass spectrum of the model was obtained in \cite{15} via a sigma
model interpretation, i.e. linear combinations of the Liouville field
$\sigma $ (see below) and the dilaton $\phi $ are considered as new string
coordinates. In the following we will call that point of view $d+2$
dimensional critical string.

Our paper is organized as follows. In the second section we calculate
correlators in the gravity part of the theory. That will be a
preparation of the third section where correlators of gravitational dressed
tachyons are written down and their gravitational dressed dimensions are
calculated. In the fourth section we consider the $2+d$ dimensional
critical string and give an expression in the form of
Shapiro--Virasoro amplitudes. A rough argumentation provides the poles
of single scattering channels. In $d=12$ dimensions the four point
function will be considered more detailed. We get poles in S,T and
U channels and also leg poles which are created by scattering with
background tachyons of fixed momentum.

\section{Correlators in 2D gravity}
\setcounter{equation}{0}
We will confine ourself to simply connected world sheets and use the
conformal gauge
\begin{equation}
g_{\alpha \beta }=e^{2\sigma (z)}\delta _{\alpha \beta }.
\end{equation}
The 2D gravity action contains besides the pure gravity action (1.1)
the matter induced Liouville action \cite{1},
\begin{equation}
S_{g}=S_{pg}+(26-d)S_{L},
\end{equation}
\begin{equation}
S_{L}=\frac{1}{12\pi }\int d^{2}z\, \left( \frac{1}{2}\partial _{\alpha }
\sigma \partial _{\alpha }\sigma +\mu e^{2\sigma }\right) ,
\end{equation}
where $\mu $ is the Liouville mass. We define fixed area expectation
values as follows,
$$ Z(A) = \int \mbox{$\cal{D}$}_{g}\phi \mbox{$\cal{D}$}_{g}\sigma
\, d\Lambda \, \delta \left( 1-\frac{1}{A}\int d^{2}z\sqrt{g}\right)
\, \delta \left( \Lambda A
+4\pi \right) \, e^{-S_{g}} $$
\begin{equation}
\langle \cdots \rangle _{\mid A}= \frac{1}{Z(A)}\int \mbox{$\cal{D}$}_{g}\phi
\mbox{$\cal{D}$}_{g}\sigma  \, d\Lambda \,
\delta \left( 1-\frac{1}{A}\int d^{2}z
\sqrt{g}\right) \, \delta \left( \Lambda A+4\pi \right) \, e^{-S_{g}}
\left( \cdots \right) .
\end{equation}
The second delta function comes from the constraint of constant curvature
and the Gau\ss --Bonnet theorem. The first delta function is redundant,
because the constraint of constant curvature implies a constant area.
However, when we use translation invariant measures the term
$\phi \sqrt{g}$ in (1.1) will be renormalized and the constraint of constant
curvature is no longer manifest. The arguments of delta functions are
arranged in such a way that $Z=\int dAZ(A)$ is dimensionless.
Now we consider the correlator of N 2D gravitons and dilatons,
$$\langle \prod _{i=1}^{N}e^{2\beta _{i}\sigma (z_{i})+2i\gamma _{i}
\phi (z_{i})} \rangle _{\mid A} .$$
In order to be able to calculate this correlator we have to move to
translation invariant measures \cite{4,5}.
Therefore we split $\sigma $ into a classical part $\hat {\sigma }$
and into a quantum part $\sigma $ and turn to measures with $g$
replaced by $\hat{g}$ ,
$$\mbox{$\cal{D}$}_{g}\longrightarrow \mbox{$\cal{D}$}_{\hat{g}},
\; \; \hat{g}_{\alpha \beta }=e^{2\hat{\sigma }}\delta _{\alpha \beta }.$$
The Jacobian can be calculated by methods given in \cite{6} or \cite{20}.
We get
\begin{eqnarray}
\lefteqn{\langle \prod _{i=1}^{N}e^{2\beta _{i}\sigma (z_{i})
+2i\gamma _{i}\phi (z_{i})}\rangle _{\mid A} =} \nonumber \\
& & \frac{1}{Z(A)}e^{-(26-d)S_{L}[\hat{\sigma }]}
\prod_{i=1}^{N}e^{2\beta _{i}\hat{\sigma }(z_{i})} \nonumber \\
& & \int \mbox{$\cal{D}$}_{\hat{g}}\sigma \mbox{$\cal{D}$}_{\hat{g}}\phi
e^{-12a\hat{S}_{L}-\hat{S}_{pg}}\int  \, d\Lambda \, \delta
\left( 1-\frac{1}{A}
\int d^{2}z \sqrt{\hat{g}}(e^{2\sigma })_{ren}\right) \nonumber \\
& &\delta \left( \Lambda A+
4\pi \right) \prod_{i=1}^{N}e^{2\beta _{i}\sigma (z_{i})+
2i\gamma _{i}\phi (z_{i})},
\end{eqnarray}
where
$$ a=\frac{24-d}{12}$$
$$\hat{S}_{L}=\frac{1}{12\pi }\int d^{2}z\sqrt{\hat{g}}\, \left( \frac{1}{2}
\hat{g}^{\alpha \beta }\partial _{\alpha }\sigma \partial _{\beta }
\sigma +\hat{R}\sigma + \mu (e^{2\sigma })_{ren}\right) $$
\begin{equation}
\hat{S}_{pg}=\frac{i}{\pi }\int d^{2}z\sqrt{\hat{g}}\, \left[ \phi
(\hat{\Delta }\sigma+
\hat{R} )+\Lambda (\phi e^{2\sigma})_{ren}\right] .
\end{equation}
It is useful to introduce
\begin{equation}
\psi =\sigma +\frac{i}{a}\phi
\end{equation}
instead of $\sigma $. Then the massless part of the action becomes diagonal,
\begin{eqnarray}
\lefteqn{12a\hat{S}_{L\, \mid \mu =0}+\hat{S}_{pg \, \mid \Lambda =0} =}
\nonumber \\
& & \frac{a}{\pi }\int d^{2}z\sqrt{\hat{g}}\, \left( \frac{1}{2}
\hat{g}^{\alpha \beta }\partial _{\alpha }\psi \partial _{\beta }\psi
+\hat{R}\psi \right) + \nonumber \\
& & \frac{1}{a\pi }\int d^{2}z\sqrt{\hat{g}}\, \frac{1}{2}
\hat{g}^{\alpha \beta }\partial _{\alpha }\phi \partial _{\beta }\phi .
\end{eqnarray}
(2.8) is a conformal field theory with central charge
$$c=12a+1+1=26-d$$
which cancels the central charge of the matter part of the theory
(compare next section). In (2.5) we have admitted a renormalization
of $e^{2\sigma }$ and $\phi e^{2\sigma }$. The operators
$(e^{2\sigma })_{ren}$ and $(\phi e^{2\sigma })_{ren}$ must not destroy the
conformal invariance and therefore we require them to be primary fields of
dimension (1,1). (Otherwise we would get $\hat{\sigma }$ dependent
expressions). The conformal dimension of a general composite
vertex operator can be calculated via an operator product expansion.
That provides
\begin{eqnarray}
\Delta \left( e^{2\rho \sigma }e^{2i\omega \phi }\right)  & = &
\Delta \left( e^{2\rho \psi }\right) +\Delta \left(
e^{2i(\omega -\frac{\rho }{a})\phi }\right)
\nonumber \\
 & = & \rho -\frac{\rho ^{2}}{2a}+\frac{a}{2}\left( \omega
-\frac{\rho }{a}\right) ^{2}.
\end{eqnarray}
The requirement $\Delta =1$ yields one equation for two parameters and hence
we have infinite many operators of dimension (1,1), (and there are even
more of them \cite{15}). A natural additional requirement is that
renormalized and unrenormalized operators should coincide in the
semi classical limit ($a\longrightarrow \infty $). That leads to
\begin{equation}
(e^{2\sigma })_{ren}=e^{2\sigma }
\end{equation}
\begin{eqnarray}
(\phi e^{2\sigma })_{ren} & = & -\frac{ia}{4}e^{2\sigma }
(e^{\frac{4i}{a}\phi }-1)= \nonumber \\
 & = & \phi e^{2\sigma }+o\left( \frac{1}{a}\right) .
\end{eqnarray}
Performing zero mode, and $\Lambda $ integration and neglecting
uninteresting factors provides
\begin{eqnarray}
\lefteqn{\langle \prod_{i=1}^{N}e^{2\beta _{i}\sigma (z_{i})+
2i\gamma _{i}\phi (z_{i})}\rangle _{\mid A}=}\nonumber \\
& & \frac{A^{-t-s-1}}{Z(A)}e^{-\mu A}\Gamma (-t) \prod_{i=1}^{N}
e^{2\beta _{i}\hat{\sigma }(z_{i})}
e^{-(26-d)S_{L}[\hat{\sigma }]} \nonumber \\
& & \int \mbox{$\cal{D}$}_{\hat{g}\perp }\sigma
\mbox{$\cal{D}$}_{\hat{g} \perp}\phi \,
e^{-\hat{S}_{pg\, \mid \Lambda =0}-\hat{S}_{L\, \mid \mu =0}}
\left( \int d^{2}z\sqrt{\hat{g}}e^{2\sigma }e^{\frac{4i}{a}\phi }\right) ^{t}
\nonumber \\
& & \left( \int d^{2}z\sqrt{\hat{g}}e^{2\sigma }\right) ^{s}
\prod_{i=1}^{N}e^{2\beta _{i}\sigma (z_{i})+2i\gamma _{i}\phi (z_{i})},
\end{eqnarray}
where
$$t=a\left( 1-\frac{1}{2}\sum_{i=1}^{N}\gamma _{i}\right) \; \; \mbox{and}$$
\begin{equation}
s=2a-t-\sum_{i=1}^{N}\beta _{i}=a +\sum_{i=1}^{N}\left( \frac{a}{2}
\gamma _{i} -\beta _{i}\right) .
\end{equation}
The right angle indicates that zero modes are integrated out.
We observe that the partition function behaves like
\begin{equation}
Z(A) \sim A^{-2a-1}e^{-\mu A}=A^{\frac{d-24}{6}-1}e^{-\mu A}
\end{equation}
which coincides with the calculation with non translation invariant measures
performed in \cite{11}.
The calculation of the partition function is rather simple with non
translation invariant measures.
In the case of general correlators it turns out to be more difficult.
For non vanishing $\gamma _{i}$'s one has to solve the Liouville
equation with additional delta function like sources as it was stated
in \cite{15}. But also for vanishing $\gamma _{i}$'s the problem is
involved, because one has to integrate over all solutions of the
Liouville equation. If one is only interested in the partition function
the integrand will not depend on the special form of the solution of the
Liouville equation and the integral over all solutions gives  an uninteresting
factor.
The situation is much more complicated for correlation functions
where the integrand depends on the special form of the solution.
However, if one was able to solve that problem one could check
whether our assumptions (2.10) and (2.11) are correct at the level of
higher point functions, too.

Now we return to (2.12) and define the zero modes in a covariant way
\cite{6}
\begin{eqnarray}
\phi _{0} & = & \frac{1}{4\pi }\int d^{2}z\sqrt{\hat{g}}\hat{R}\phi
\nonumber \\
\sigma _{0} & = & \frac{1}{4\pi }\int d^{2}z\sqrt{\hat{g}}\hat{R}\sigma .
\end{eqnarray}
Then we use the field redefinition (2.7) and get, up to a factor,
for $\psi $ and $\phi $ the same $\hat{\sigma }$--dependent
propagator
\begin{equation}
G(z,z\prime \mid \hat{\sigma })= -\log (M\mid z-z\prime \mid)-
\frac{\hat{\sigma }(z)}{2}-\frac{\hat{\sigma }(z\prime )}{2}+
3S_{L}[\hat{\sigma }].
\end{equation}
As usual (compare \underline{e.g.} \cite{16}) we take $t$ and $s$
as positive integers during the calculation and assume that final
results could be continued to real values. That leads to
\begin{eqnarray}
\lefteqn{\langle \prod_{i=1}^{N}e^{2\beta _{i}\sigma (z_{i})+
2i\gamma _{i}\phi (z_{i})}\rangle _{\mid A}=}\nonumber \\
& & \frac{\Gamma (-t)}{\Gamma (-a)}A^{2a-t-s}
e^{-12aS_{L}[\hat{\sigma }]}\prod_{i=1}^{N}e^{2\beta _{i}\hat{\sigma }(z_{i})}
\nonumber \\
& & \int
\left( \prod_{i=N+1}^{N+t+s}d^{2}z_{i}\, e^{2\beta _{i}\hat{\sigma }
(z_{i})}\right)
\prod_{i,j}^{N+t+s}
e^{(2\gamma _{i}\beta _{j}-a\gamma _{i}\gamma _{j})G(z_{i},z_{j} \mid
\hat{\sigma })},
\end{eqnarray}
where
$$\gamma _{N+1}=\cdots =\gamma _{N+t}=\frac{2}{a}$$
\begin{equation}
\gamma _{N+t+1}=\cdots =\gamma _{N+t+s}=0
\end{equation}
$$\beta _{N+1}=\cdots =\beta _{N+t+s}=1$$
which implies
\begin{equation}
\sum_{i=1}^{N+t+s}\gamma _{i}=2 \; \; \mbox{and} \; \;
\sum_{i=1}^{N+t+s}\beta _{i}=2a.
\end{equation}
It is easy to see that the $\hat{\sigma }$ dependence drops out which
represents a nontrivial check of our calculation.

We regularize $\log (M0)$ by $\log(M\epsilon /\eta )$, $\eta $ is a
renormalization scale and $\epsilon $ is a UV cut off. Finally we get
\begin{eqnarray}
\lefteqn{\langle \prod_{i=1}^{N}e^{2\beta _{i}\sigma (z_{i})+
2i\gamma _{i}\phi (z_{i})}\rangle _{\mid A}=}\nonumber \\
& & A^{2a-t-s}\frac{\Gamma (-t)}{\Gamma (-a)}
\left( \frac{M\epsilon }{\eta }\right) ^{-\sum_{i=1}^{N}(2\gamma _{i}\beta _{i}
-\gamma _{i}^{2}a)} \nonumber \\
& & \int \prod_{i=1}^{t}d^{2}u_{i}\prod_{l=1}^{s}d^{2}w_{l}
\prod_{j=1}^{N}\prod_{i=1}^{t}(M\mid z_{j}-u_{i}\mid )
^{-\frac{4}{a}\beta _{j}+2\gamma _{j}}
\prod_{j=1}^{N}\prod_{l=1}^{s}(M\mid z_{j}-w_{l}\mid )^{-2\gamma _{j}}
\nonumber \\
& & \prod_{l=1}^{s}\prod_{i=1}^{t}(M\mid u_{i}-w_{l}\mid )^{-\frac{4}{a}}
\prod_{i\not= j}^{N}(M\mid z_{i}-z_{j}\mid )
^{\gamma _{i}\gamma _{j}a-2\gamma _{i}\beta _{j}}.
\end{eqnarray}

\section{Matter coupled to 2D gravity}
\setcounter{equation}{0}
In the conformal gauge (2.1) the string action is given by
\begin{equation}
S_{M}=\frac{1}{4\pi }\int d^{2}z\frac{1}{2}\partial _{\alpha }X^{\mu }(z)
\partial _{\alpha }X_{\mu }(z), \; \; \mu =1,\cdots ,d.
\end{equation}
Like the authors of \cite{6} we define a normal ordered tachyon operator
by multiplying it with a conformal cut off and a scalar density. We
allow that 2D dilatons take part in the scalar density,
\begin{equation}
:e^{ik_{j}\cdot X(z_{j})}: = e^{ik_{j}\cdot X(z_{j})}B_{j}(z_{j})
\end{equation}
where
\begin{equation}
B_{i}(z_{i})=\left( \epsilon ^{2}e^{2\sigma (z_{i})}\right) ^{\beta _{i}-1}
e^{2\sigma (z_{i})}\, e^{2i\gamma _{i}\phi (z_{i})}.
\end{equation}
The N-point tachyon correlator is then given by
\begin{eqnarray}
\lefteqn{ \langle  \prod_{i=1}^{N} :e^{ik_{i}\cdot X(z_{i})}:\rangle =}
\nonumber \\
& & \frac{1}{Z} \int \mbox{$\cal{D}$}_{g}\sigma\mbox{$\cal{D}$}_{g}\phi
\mbox{$\cal{D}$}_{g}X\mbox{$\cal{D}$}_{g}(ghost)
e^{-S_{pg}-S_{M}-S_{ghost}}\prod_{i=1}^{N}:e^{ik_{i}\cdot X(z_{i})}:,
\end{eqnarray}
where the ghost integral arises from conformal gauge fixing.
Using the result (2.20) from the previous section we get
\begin{eqnarray}
\lefteqn{ \langle \prod_{i=1}^{N}:e^{ik_{i}\cdot X(z_{i})}:\rangle _{\mid A}=}
\nonumber \\
& & A^{2a-t-s}\frac{\Gamma (-t)}{\Gamma (-a)}
\delta ^{(d)}\left( \sum_{i=1}^{N}k_{i} \right)
\left( \frac{\epsilon M}{\eta }\right) ^{\sum_{i=1}^{N}
(k_{i}^{2}-2\gamma _{i}\beta _{i}+\gamma _{i}^{2}a)}
\epsilon ^{2\sum_{i=1}^{N}(\beta _{i}-1)}\nonumber \\
& & \int \prod_{i=1}^{t}d^{2}u_{i}\prod_{l=1}^{s}d^{2}w_{l}\prod_{j=1}^{N}
\prod_{i=1}^{t}(M\mid z_{j}-u_{i}\mid )^{-\frac{4}{a}\beta _{j}+2\gamma _{j}}
\prod_{j=1}^{N}\prod_{l=1}^{s}(M\mid z_{j}-w_{l}\mid )^{-2\gamma _{j}}
\nonumber \\
& & \prod_{l=1}^{s}\prod_{i=1}^{t}(M\mid u_{i}-w_{l}\mid )^{-\frac{4}{a}}
\prod_{i \not= j}^{N}(M\mid z_{i}-z_{j}\mid )^{k_{i}\cdot k_{j}+
\gamma _{i}\gamma _{j}a-2\gamma _{i}\beta _{j}}.
\end{eqnarray}
Requiring independence on the UV cut off $\epsilon $ yields
\begin{equation}
\beta _{i}-\gamma _{i}\beta _{i}+\frac{\gamma _{i}^{2}}{2}a
=1-\frac{k_{i}^{2}}{2}\equiv 1-\Delta _{i}^{(0)}
\end{equation}
where $\Delta ^{(0)}_{i}$ is the bare conformal dimension, i.e. the
conformal dimension with respect to the matter part only. With (2.9)
and (3.6) follows
\begin{equation}
\Delta \left( B_{i}(z_{i})\right) =1-\Delta _{i}^{(0)}.
\end{equation}
Thus our definition of gravitational dressing is equivalent to the definition
used in \cite{5}. The gravitational dressed dimensions are defined via the
scaling behavior of (3.5),
\begin{equation}
\langle \prod_{i=1}^{N}:e^{k_{i}\cdot X(z_{i})}:\rangle _{\mid A}\sim
\prod_{i=1}^{N}A^{1-\Delta _{i}}.
\end{equation}
That leads to
\begin{equation}
\Delta _{i}=1-\beta _{i}
\end{equation}
and is not unique due to the presence of the $\gamma _{i}$.
The most natural restriction is
\begin{equation}
\gamma _{i}=0,
\end{equation}
i.e. gravitational dressing is carried by gravitons only. It ensures also
that the unit operator has scaling dimension zero. Restriction (3.10)
leads to a trivial KPZ relation
\begin{equation}
\Delta _{i}=\Delta _{i}^{(0)}.
\end{equation}
(3.11) coincides with the perturbative result \cite{12}.

\section{$d+2$ dimensional critical string}
\setcounter{equation}{0}
We consider the integrated N point functions
\begin{equation}
A_{N}(k_{1},\cdots ,k_{N})=\frac{1}{Vol(SL(2,R))}\int \prod_{i=1}^{N}
d^{2}z_{i}\langle \prod_{i=1}^{N}:e^{ik_{i}\cdot X(z_{i})}:\rangle
\end{equation}
and rewrite them in a suggestive form, (we neglect pre factors),
\begin{eqnarray}
\lefteqn{A_{N}(k_{1},\cdots ,k_{N})=} \nonumber \\
& & \frac{1}{Vol(SL(2,R))}\int \prod_{j=1}^{N}d^{2}z_{j}
\prod_{\alpha =1}^{t}d^{2}u_{\alpha }
\prod_{J=1}^{s}d^{2}w_{J} \nonumber \\
& & \prod_{i<j}\mid z_{i}-z_{j}\mid ^{2K_{i}\cdot K_{j}}
\prod_{j,\alpha }\mid z_{j}-u_{\alpha }\mid ^{2iK_{j}\cdot \bar{K}}\nonumber \\
& & \prod_{j,J}\mid z_{j}-w_{J}\mid ^{2iK_{j}\cdot K}
\prod_{\alpha ,J}\mid u_{\alpha }-w_{J}\mid ^{-2K\cdot \bar{K}},
\end{eqnarray}
where we have introduced the index conventions
$$j=1,\ldots ,N;\; \; \alpha =1,\ldots ,t;\; \; J=1,\ldots ,s;$$
and the $K_{i}$ are $d+2$ dimensional vectors. Comparison of (4.2) with
(3.5) provides
\begin{equation}
K_{j}=\left( k_{j1},\cdots ,k_{jd},\frac{i\beta _{j}}{\sqrt{a}},
\sqrt{a}\gamma _{j}-\frac{\beta _{j}}{\sqrt{a}}\right),
\end{equation}
\begin{equation}
i\bar{K}=\left( \underbrace{0,\cdots
,0}_{d}\frac{i}{\sqrt{a}},\frac{1}{\sqrt{a}}
\right)
\end{equation}
\begin{equation}
iK=\left( \underbrace{0,\cdots ,0}_{d},\frac{i}{\sqrt{a}},-\frac{1}{\sqrt{a}}
\right).
\end{equation}
Equation (3.6) can be written in the form
\begin{equation}
-m_{i}^{2}\equiv K_{i}^{2}+n\cdot K_{i}=2,
\end{equation}
with
\begin{equation}
n=(\underbrace{0,\cdots ,0}_{d},-2i\sqrt{a},0)
\end{equation}
(4.6) is the mass shell condition of \cite{15}.
The scalar product in (4.4) is Euclidian and the time like coordinate
is the pure imaginary one. (The case $a=0$ is the ordinary 26 dimensional
critical string and is not considered here). However, (4.6) could as well
be written in the form
$$ -\tilde{m}_{i}^{2}\equiv (K_{i}+\frac{n}{2})^{2}=2+\frac{n^{2}}{4}=2-a.$$
Then the `tachyon' would become massless in $0+2$ dimensions which is
common for two dimensional critical strings \cite{21}. In the following
we will use the mass definition (4.6).

One can extract the poles of a single channel by splitting the integration
intervals and considering the region where $\mid z_{i}-z_{j}\mid $ is
smaller than all other distances \cite{17}. Than one expands the integrand
in a Taylor series around $\mid z_{i}-z_{j}\mid =0$ and integrates
out that relative distance. That leads to the following poles,
\begin{equation}
S_{kl}\equiv (K_{k}+K_{l})^{2}+n\cdot (K_{k}+K_{l})=2-2j
\end{equation}
\begin{equation}
S_{kI}\equiv (K_{k}+i\bar{K})^{2}+n\cdot (K_{k}+i\bar{K})=2-2j
\end{equation}
\begin{equation}
S_{k\alpha }\equiv (K_{k}+iK)^{2}+n\cdot (K_{k}+iK)=2-2j
\end{equation}
\begin{equation}
S_{I\alpha }\equiv (iK+i\bar{K})^{2}+n\cdot (iK+i\bar{K})=2-2j
\end{equation}
where $j$ is a non negative integer. From (4.8) we get the same mass
spectrum as the author of \cite{15}.
Insertion of (4.4) and (4.5) provides
\begin{equation}
S_{kI}=4-\frac{4\beta _{k}}{a}+2\gamma _{k}
\end{equation}
\begin{equation}
S_{k\alpha }=4-2\gamma _{k}
\end{equation}
\begin{equation}
S_{I\alpha }=4-\frac{4}{a}
\end{equation}
Due to the presence of background tachyons with fixed momenta we have
leg poles (4.12) and (4.13). For
\begin{equation}
a=\frac{4}{2j+2},
\end{equation}
($j=$ non negative integer), we expect divergent expressions because we are
then exactly at poles of $(I,\alpha )$-channels. These divergencies can
be regularized by a cut off $\mid u_{\alpha }-w_{J}\mid >\lambda$.

Now we want to consider the four point function in more detail.
Because we have not succeeded in the continuation to real
$t$ and $s$ until now we restrict ourself to the case $a=1$ (i.e. $d=12$).
Here we remark that in the super symmetric case one gets very similar
formulas with 2D super space integrals and $a=(8-d)/4$. Then $a=1$
corresponds to the physically interesting case $d=4$. A detailed discussion
of the super symmetric case will be given in \cite{18}.
Furthermore we set
\begin{equation}
\gamma _{1}=\gamma _{2}=\gamma _{3}=\gamma _{4}=0,
\end{equation}
i.e. gravitational dressing is carried by gravitons only. Thus we have
$t=1$ and $s=1-\sum \beta_{i}$ and do not need the continuation to
real values of $t$. We use M\"obius invariance to set
\begin{equation}
z_{1}=0,\, z_{2}=1,\, z_{3}=\infty ,\, z_{4}=z
\end{equation}
and get
\begin{eqnarray}
\lefteqn{A_{4}(k_{1},k_{2},k_{3},k_{4})=}\nonumber \\
& & \int d^{2}z\int d^{2}w \int \prod_{\alpha =1}^{s}d^{2}u_{\alpha }\,
\mid z\mid ^{2k_{1}k_{4}}\mid 1-z\mid ^{2k_{2}k_{4}}
\mid z-w\mid ^{-4\beta _{4}}\nonumber \\
& & \mid w\mid ^{-4\beta _{1}}\mid 1-w\mid ^{-4\beta _{2}}
\prod_{\alpha =1}^{s}\mid u_{\alpha }-w\mid ^{-4}.
\end{eqnarray}
The calculation of the $z$-integral is given in \cite{19}.
There the two dimensional integral is expressed in terms of contour integrals
which provide the hypergeometric function. Before giving the result we
introduce suitable Mandelstam variables,
$$ S= (k_{3}+k_{4})^{2}+2(\beta _{3}+\beta _{4})$$
$$T=(k_{1}+k_{4})^{2}+2(\beta _{1}+\beta _{4})$$
\begin{equation}
U=(k_{2}+k_{4})^{2}+2(\beta _{2}+\beta _{4}).
\end{equation}
Adding the Mandelstam variables provides
\begin{equation}
S+T+U=8+4\beta  _{4}=-4m^{2}+4\beta _{4},
\end{equation}
where $m^{2}=-2$ is the tachyon mass. The sum (4.20) is not constant
because background tachyons take part in scattering. With the help of
\begin{equation}
K_{5}=(\underbrace{0,\cdots ,0}_{d},i,1)=i\bar{K}
\end{equation}
we define a further Mandelstam variable ($\beta _{5}=1$)
\begin{equation}
V=(K_{4}+K_{5})^{2}+2(\beta _{4}+\beta _{5})
\end{equation}
and get
\begin{equation}
S+T+U+V=12=-6m^{2}.
\end{equation}
We note that (4.16) implies that there is no interaction between external
tachyons and the $s$ background tachyons
$$\left( \int d^{2}z\, e^{2\sigma}\right) ^{s}.$$
In terms of $S,T,U,$ and $V$ (4.17) becomes
\begin{eqnarray}
\lefteqn{A_{4}= }\nonumber \\
& & \pi \int d^{2}w\{
\frac{\Delta (\frac{1}{2}T-1)\Delta (\frac{1}{2}V-1)}{\Delta (\frac{1}{2}
(T+V)-2)}\mid w\mid ^{T+V-4}\mid F(2-\frac{1}{2}U,\frac{1}{2}T-1,
\frac{1}{2}(T+V)-2;w)\mid ^{2}+\nonumber \\
& & \frac{\Delta (\frac{1}{2}S-1)\Delta(\frac{1}{2}U-1)}{\Delta
(\frac{1}{2}(U+S)-2)}\mid F(2-\frac{1}{2}V,\frac{1}{2}S-1,\frac{1}{2}(U+S)-2;
w)\mid^{2}\} \nonumber \\
& & \mid w\mid ^{-4\beta _{1}}\mid 1-w\mid ^{-4\beta _{2}}
\int \prod_{\alpha =1}^{s}d^{2}u_{\alpha }\mid u_{\alpha }-w\mid ^{-4}.
\end{eqnarray}
$F$ is the hypergeometric function and
$$\Delta (x)=\frac{\Gamma (x)}{\Gamma (1-x)}.$$
In the S-channel for example we observe the poles
\begin{equation}
S=2-2j
\end{equation}
which confirms (4.8). In fact poles in $V$ are leg poles because
\begin{equation}
V=4-4\beta _{4}.
\end{equation}
These leg poles occur because background tachyons with fixed momenta take
part in scattering. The $u$-integrals are divergent because $a=1$ is
contained in (4.15). The divergence can be regularized by a cut off
$$\mid u_{\alpha }-w\mid >\lambda$$
and renormalized via
$$\left( \int d^{2}ze^{2\sigma }\right) ^{s}\longrightarrow \left(
\lambda ^{2}\int d^{2}z  e^{2\sigma }\right)^{s}.$$
Then one can solve the $u$-integrals. Expanding the $w$-integrand
in a series around $\mid w\mid =0$ and $\mid w\mid =\infty $
one can convince himself that there are no additional poles in
S,T, and U channels.

One could regard (4.24) also as an off shell 12 dimensional non critical
string amplitude. That would be closer to the original approach of sections
1, 2, and 3. Then poles in scattering amplitudes occur when the
intermediate particles are gravitationally dressed according to equation (3.6).
Unfortunately we do not know how to get a mass spectrum in such a picture.
However, we can guess a mass spectrum by consistency requirements.
Suppose we are given a ground state or `tachyon' mass $m_{0}$.
Then the on shell four point function should have a pole at
\begin{equation}
(k_{3}+k_{4})^{2}=-m_{0}^{2}
\end{equation}
Together with equation (4.25) that leads to the restriction
\begin{equation}
m_{0}^{2}=-2j_{0}-2
\end{equation}
with $j_{0}$ a non negative integer. The on shell value of $\beta _{i}$
is then
$$\beta _{i}=1+\frac{m_{0}^{2}}{2}=-j_{0}.$$
Hence there are no additional divergences of the on shell amplitude
due to leg poles. We obtain the full mass spectrum via (4.25)
$$2-2j=-M_{j}^{2}+4+2m_{0}^{2}$$
\begin{equation}
M_{j}^{2}=2j-2-4j_{0},\; \; j=0,1,2,\ldots
\end{equation}
Now we require the ground state to be the lightest one,
\begin{equation}
M_{j}^{2}\geq m_{0}^{2},\;  \; \forall j.
\end{equation}
That leads to
\begin{equation}
j_{0}=0
\end{equation}
and we obtain the same mass spectrum as in 26 dimensional critical string
theory.

\section{Conclusions}
\setcounter{equation}{0}
We have calculated the N point tachyon amplitude in a model where the
gravitational part is trivialized by the constraint of constant
curvature. Reasonable assumptions provided the correct string
susceptibility and a trivial KPZ relation.

Although we were not able to give a closed formula for arbitrary
dimensions and external momenta we got an expression for the four
point function where poles in different scattering channels
are manifest. Interpreting the model as $d+2$ dimensional critical
string we obtained, like the author of \cite{15}, the same mass
spectrum as in 26 dimensional critical string theory. Furthermore
we have obtained hints that also a $d$ dimensional non critical
string picture provides the same mass spectrum.

The most serious open problem is the continuation to real values of
$t$ and $s$. The upper barrier $d=1$  of the model with induced
gravity only is in the model considered here a lower barrier at
$d=0$ ($a=2$). We have doubts whether one can use our
calculation directly for $d=0$. In the gravity part of our theory
M\"obius invariance is not manifest, e.g. the two point function
is not zero for different conformal weights. We can hope that the zero
mode integration of the matter part selects configurations where
M\"obius invariance survives.
But for $d=0$ there is no matter part. Therefore we expect
that the lower barrier, $d=0$, should be considered as
a limit $c_{matter}\longrightarrow 0$. In order to be able to
perform that limit one needs the continuation to real values
of $t$ and $s$ urgently.
\\
\\
\begin{center}
\bf{Acknoledgements}
\end{center}
I am very grateful to H.-J. Otto for excellent explanations and many
detailed hints.\\
Furthermore I would like to thank H. Dorn and K. Behrndt for useful
discussions.


\begin{thebibliography}{20}
\bibitem{1}A.M. Polyakov, Phys. Lett. {\bf B113}(1981)207
\bibitem{2}A.M. Polyakov, Mod. Phys. Lett. {\bf A2}(1987)899
\bibitem{3}V.G. Knizhnik, A.M. Polyakov, A.A. Zamolodchikov,
Mod. Phys. Lett. {\bf A3}(1988)819
\bibitem{4}F. David, Mod. Phys. Lett. {\bf A3}(1988)1651
\bibitem{5}J. Distler, H. Kawai, Nucl. Phys. {\bf B321}(1989)509
\bibitem{6}H. Dorn, H.-J. Otto, Phys. Lett. {\bf B232}(1989)327;
 CERN-TH. 6285/91
\bibitem{7}R. Jackiw, Quantum theory of gravity (ed. S. Christiensen,
Adam Hilger, Bristol 1984); p. 403\\
C. Teitelboim, Quantum theory of gravity; p. 327
\bibitem{8}A.H. Chamseddine, D. Wyler, Nucl. Phys. {\bf B340}(1990)595
\bibitem{9}E. D'Hoker, D.H. Phong, Rev. Mod. Phys. {\bf 60}(1988)987
\bibitem{10}I.M. Lichtzier, S.D. Odintsov, Mod. Phys. Lett. {\bf A6}(1991)1953
\bibitem{11}A.H. Chamseddine, Phys. Lett. {\bf 256B}(1991)379;
{\bf 258B}(1991)97
\bibitem{12}F.D. Mazzitelli, N. Mohammedi, Phys. Lett. {\bf B268}(1991)12
\bibitem{13}S. F\"orste, Proceedings XXV Int. Symp. Ahrenshoop (Sept. 1991)
\bibitem{14}S.D. Odintsov, I.L. Shapiro, Madrid preprint(1991), FTUAM-91-33
\bibitem{15}A.H. Chamseddine, Nucl. Phys. {\bf B368}(1992)98
\bibitem{20}N. Mavromatos, J. Miramontes, Mod. Phys. Lett.
{\bf A4}(1989)1849;\\
E. D'Hoker, P.S. Kurzepa, Mod. Phys. Lett. {\bf A5}(1990)1411
\bibitem{16}M. Goulian, M. Li, Phys. Rev. Lett. {\bf 66}(1991)2051
\bibitem{21}D. Kutasov: Some properties of (non)critical strings, PUPT-1277
\bibitem{17}J.A. Shapiro, Phys. Lett. {\bf 33B}(1970)361
\bibitem{18}S. F\"orste, in preparation
\bibitem{19}Vl. S. Dotsenko: Lectures on Conformal Field Theory
(Kyoto Univ., Oct. 1986)
\end{thebibliography}
\end{document}